\documentclass[twocolumn]{aastex61}

\newcommand\aastex{AAS\TeX}

\received{2018 June 25}
\revised{2018 August 16}
\accepted{2018 August 16}

\shorttitle{\aastex\ sample article}
\shortauthors{Jiang et al.}

\begin{document}

\title{Surface Radioactivity or Interactions? Multiple Origins of Early-excess Type Ia Supernovae and Associated Subclasses}

\correspondingauthor{Ji-an Jiang}
\email{yuzhoujiang@ioa.s.u-tokyo.ac.jp}

\author{Ji-an Jiang}
\affil{Institute of Astronomy, Graduate School of Science, The University of Tokyo, 2-21-1 Osawa, Mitaka, Tokyo 181-0015, Japan}
\affil{Department of Astronomy, Graduate School of Science, The University of Tokyo, 7-3-1 Hongo, Bunkyo-ku, Tokyo 113-0033, Japan}

\author{Mamoru Doi}
\affil{Institute of Astronomy, Graduate School of Science, The University of Tokyo, 2-21-1 Osawa, Mitaka, Tokyo 181-0015, Japan}
\affil{Research Center for the Early Universe, Graduate School of Science, The University of Tokyo, 7-3-1 Hongo, Bunkyo-ku, Tokyo 113-0033, Japan}
\affil{Kavli Institute for the Physics and Mathematics of the Universe (WPI), The University of Tokyo, 5-1-5 Kashiwanoha, Kashiwa, Chiba 277-8583, Japan}

\author{Keiichi Maeda}
\affil{Department of Astronomy, Kyoto University, Kitashirakawa-Oiwake-cho, Sakyo-ku, Kyoto 606-8502, Japan}
\affil{Kavli Institute for the Physics and Mathematics of the Universe (WPI), The University of Tokyo, 5-1-5 Kashiwanoha, Kashiwa, Chiba 277-8583, Japan}

\author{Toshikazu Shigeyama}
\affil{Research Center for the Early Universe, Graduate School of Science, The University of Tokyo, 7-3-1 Hongo, Bunkyo-ku, Tokyo 113-0033, Japan}



\begin{abstract}

Early-phase Type Ia supernovae (SNe Ia), especially those with luminosity enhancement within the first few days of explosions (``early-excess SNe Ia"), play an irreplaceable role in addressing the long-standing progenitor and explosion issue of SNe Ia. In this paper, we systematically investigate 11 early-excess SNe Ia from subluminous to luminous subclasses. Eight of them are selected from 23 SNe Ia with extremely early-phase optical light curves (``golden" early-phase SNe Ia), and three of them are selected from 40 SNe Ia (including 14 golden samples) with early-phase UV/NUV light curves. We found that previously discovered early-excess SNe Ia show a clear preference for specific SN Ia subclasses. In particular, the early-excess feature shown in all six luminous (91T- and 99aa-like) SNe Ia is in conflict with the viewing angle dependence predicted by the companion-ejecta interaction scenario. Instead, such a high early-excess fraction is likely related to the explosion physics of luminous SNe Ia; i.e. a more efficient detonation happening in the progenitor of luminous SNe Ia may consequently account for the early-excess feature powered by the radiation from a $^{56}$Ni-abundant outer layer. The diversity of early-excess features shown in different SN Ia subclasses suggests multiple origins of the discovered early-excess SNe Ia, challenging their applicability as a robust progenitor indicator. Further understanding of the early-excess diversity relies not only on multiband photometry and prompt-response spectroscopy of individual early-excess SNe Ia but also on investigations of the general early-phase light-curve behavior of each SN Ia subclass, which can be realized through ongoing/forthcoming transient survey projects in the near future.

\end{abstract}




\section{Introduction} \label{sec:intro}

Type Ia supernovae (SNe Ia) have been thought to originate from the thermonuclear explosion of a carbon-oxygen white dwarf (WD) in a binary system. Even though great success was achieved by using them as a cosmic distance indicator in the 1990s \citep{perlmutter97,perlmutter99,riess98}, the progenitor and the physics leading to the explosion are still under debate \citep{hillebrandt13,maoz14,Maeda16}. In the last decades, a tremendous number of SNe Ia have been found through various kinds of transient surveys, and a growing number of SNe Ia discovered within a few days of their explosions provide unique information about the progenitor system and the explosion mechanism of SNe Ia \citep{nugent11,foley11,cao15,shappee16,marion16, hosseinzadeh17,JJA2017}.

\citet{kasen10} proposed that a prominent brightening in the first few days of the explosion can be observed under specific viewing directions due to the interaction between the expanding ejecta and a nondegenerate companion star, which makes SNe Ia with additional luminosity enhancement in the early time (``early-excess SNe Ia"; EExSNe Ia) a powerful indicator for the single-degenerate (SD) progenitor system \citep{kasen10,maeda14,kutsuna15}. Since then, surveys for EExSNe Ia have become particularly popular in time-domain astronomy. The first reported EExSN Ia, iPTF14atg, has been suggested as strong evidence for the companion-ejecta interaction \citep{cao15}. Soon after, the companion-interaction scenario was also proposed as the likely origin of early-excess features of SN~2012cg and the recently discovered SN~2017cbv \citep{marion16,hosseinzadeh17}.

However, whether the early light-curve excesses of these SNe Ia are exclusively attributed to the companion interaction is still under debate. Theoretically, the interaction between dense circumstellar matter (CSM) and SN ejecta (\citealp{shen12,levanon15}, \citeyear{levanon17}; \citealp{tanikawa15,piro16,maeda18}) and vigorous mixing of radioactive $^{56}$Ni in the outermost region of SN ejecta \citep{piro16} may produce a similar early-excess feature to that predicted by the companion interaction. For instance, \citet{kromer16} claimed that the spectra of iPTF14atg cannot be reproduced by the SD scenario but show good consistency with the prediction of merging two sub-Chandrasekhar-mass WDs. If the progenitor system of iPTF14atg is not an SD system, then the early light-curve excess of iPTF14atg cannot be attributed to the companion-interaction scenario, and other effects should come into play. Different physical mechanisms have also been proposed to explain the early excess shown in SN~2012cg and SN~2017cbv \citep{shappee18,hosseinzadeh17,miller18}. Other than these ``companion-interaction-like" EExSNe Ia, the early light-curve excesses of several SNe Ia have been initially explained by different scenarios (e.g. iPTF14bdn, \citealt{smitka15}; iPTF16abc, \citealt{miller18}). In particular, \citet{JJA2017} reported a normal-brightness EExSN Ia, MUSSES1604D (SN~2016jhr), that shows a prominent but red early excess and strong Ti \textsc{ii} absorptions around the $B$-band maximum. The early excess of this peculiar object can be well explained by the radiation emitted from short-lived radioactive elements that were generated by a precursory detonation at a thin helium shell of its primary WD, which is the first robust evidence of the multiple origins of EExSNe Ia.

As opposed to the EEx discoveries in several SN Ia subclasses, there is no clear evidence of EEx detection in normal SNe Ia so far. For example, a smooth-rising light curve of SN~2011fe from the very beginning (with brightness $\sim$1000 times fainter than its peak) provides a significant constraint on the companion type \citep{nugent11,li11}. The nondetection of EEx in normal SNe Ia can be explained by either an intrinsically faint EEx due to a small nondegenerate companion star or an unfavorable viewing angle by chance under the companion-interaction scenario. However, such an observational fact is in contrast to the high EEx fraction\footnote{For SNe Ia in each subclass, an EEx fraction is defined as the number fraction of EExSNe Ia over EExSNe Ia and well-observed early-phase SNe Ia that do not show EEx features (non-EExSNe Ia; selected from papers and public resources) in the same subclass.} discovered in 91T/99aa-like SNe Ia, which may suggest differences in their natures between some SN Ia subclasses (see \S4.2).

With the largest sample of early-phase SNe Ia discovered until 2018, we present new evidence to further prove multiple origins of EExSNe Ia and discuss the explosion mechanism and progenitor of specific SN Ia subclasses from the EEx perspective. This paper is organized as follows. General information on 23 well-observed early-phase SNe Ia (``golden early-phase SNe Ia") from the subluminous to luminous subclass are summarized in \S2, and further investigations of previously reported and unnoticed EExSNe Ia are shown in \S3. Discussions about the multiple origins of EEx and their associated subclasses are given in \S4, and our conclusions are summarized in \S5.

\section{Golden early-phase SNe Ia} \label{sec:Golden early-phase SNe Ia}

In this section, we systematically investigate 23 ``golden early-phase SNe Ia" from published papers. In general, we define spectroscopically classified SNe Ia with single-band or multiband photometry taken from $\sim$16 days before to 15 days after the $B$-band maximum (in the rest frame) as golden early-phase SNe Ia. Given that some SNe Ia have intrinsically short rising phases, An SN Ia discovered at $\sim$15 days before the maximum but with a large photometric depth is also denoted as a golden sample. Here this ``large photometric depth" is quantitatively defined in a way that a magnitude difference between the detection and the maximum in the same band is larger than 4 mag (LSQ12hxx, LSQ13ry, iPTF13ebh, and SN~2016coj), or a rising rate derived from the earliest photometry and the nondetection limiting magnitude is higher than about 1 mag day$^{-1}$ (ASASSN-14lp). Note that SN~2014J (observations from early-phase to post-maximum in the same band are missing; \citealp{goobar14}) and two early-phase SNe Ia discovered by the $Kepler~Space~Telescope$, KSN~2012a and KSN~2011b (without spectral information; \citealp{olling15}), do not meet the criteria of golden early-phase SNe Ia.

All golden early-phase SNe Ia are further classified into ``normal", ``luminous," ``transitional," ``subluminous," and ``hybrid" (or ``He-det") subclasses depending on their general photometric and spectroscopic characteristics. Their basic information is listed in Table 1. We adopt ${\Delta}m_{15}(B)$ value and $B$-band apparent magnitudes at peak ($m_{B,max}$) from relevant publications. Absolute $B$-band magnitudes at peak ($M_{B,max}$) are derived by adopting a Hubble constant $H_{0}$ = 70 km s$^{-1}$ Mpc$^{-1}$ and Galactic and host extinction values from \citet{schlegel98} and relevant publications, respectively. $M_{B,max}$ are plotted against ${\Delta}m_{15}(B)$ in Figure 1.

\begin{deluxetable*}{ccccccccc}
\tablenum{1}
\tablecaption{Characteristics of Golden Early-phase SNe Ia\label{tab:messier}}
\tablewidth{0pt}
\tablehead{
\colhead{Name$^a$} & \colhead{Subclass} & \colhead{${\Delta}m_{15}(B)$} & \colhead{Peak} & \colhead{Redshift} & \colhead{Discovery} & \colhead{Photometric} & \colhead{Early Excess?} & \colhead{Data}\\
\colhead{} & \colhead{} & \colhead{} & \colhead{Magnitude$^b$} & \colhead{} & \colhead{Phase (days)$^c$} & \colhead{Depth (mag)$^d$} & \colhead{Optical/UV} & \colhead{Sources$^e$}}
\startdata
SN~2009ig & Normal & 0.89 & $-$19.07 & 0.008797 & $-$16.4 (Unf) & 4.0 & N/N & (1) \\ 
SN~2010jn & Normal & 0.90 & $-$19.49 & 0.02602 & $-$15.9 ($R48$) & 3.4 & N/-- & (2)\\ 
SN~2011fe & Normal & 1.10 & $-$19.15 & 0.000804 & $-$17.3 ($g$) & 7.3 & N/N & (3), (4)\\ 
SN~2012fr & Normal & 0.85 & $-$19.45 & 0.005453 & $-$15.9 (Unf) & 3.9 & N/N & (5), (6)\\
LSQ12hxx & Normal & 1.01 & $-$19.34 & 0.069 & $-$15.6 ($gr$) & 4.4 & N/-- & (7), (8)\\
LSQ12hzj & Normal & 1.16 & $-$19.31 & 0.029 & $-$15.9 ($gr$) & 3.7 & N/-- & (7), (8)\\
LSQ13ry & Normal & 1.27 & $-$19.46 & 0.030 & $-$15.7 ($gr$) & 4.5 & N/-- & (7), (8)\\
iPTF13dge & Normal & 1.18 & $-$19.19 & 0.0159 & $-$18.2 ($R48$) & 4.1 & N/N & (9)\\
SN~2013dy & Normal & 0.92 & $-$18.93 & 0.003889 & $-$17.6 (Unf) & 5.9 & N/-- &  (10), (11)\\
ASASSN-14lp & Normal & 0.80 & $-$19.48 & 0.005101 & $-$14.6 ($V$) & 3.5 & N/N & (12)\\
SN~2015F & Normal & 1.35 & $-$19.36 & 0.004846 & $-$16.3 ($V$) & 4.8 & N/N & (13)\\
SN~2016coj & Normal & 1.32 & $-$18.81 & 0.004523 & $-$15.4 (Unf) & 5.1 & N/-- & (14), (15)\\ \hline
\textbf{iPTF14bdn} & Luminous/91T-like & 0.84 & $-$19.44 & 0.01558 & $-$18.0 ($R48$) & 4.1 & Y/Y & (16)\\
\textbf{SN~2012cg} & Luminous/99aa-like & 0.86 & $-$19.64 & 0.00162 & $-$17.1 ($B$) & 5.2 & Y/Y & (17)\\
\textbf{iPTF16abc} & Luminous/99aa-like & 0.95 & $-$19.56 & 0.0234 & $-$17.8 ($g$) & 5.5 & Y/Y & (18)\\
\textbf{SN~2017cbv} & Luminous/99aa-like & 1.06 & $-$20.11 & 0.00399 & $-$18.9 ($B$) & 4.3 & Y/Y & (19)\\
\hline
iPTF13asv & Luminous/peculiar & 1.03 & $-$19.93 & 0.036 & $-$16.2 ($R48$) & 3.7 & N/-- & (20)\\
\textbf{LSQ12gpw} & Luminous/peculiar & 0.76 & $-$19.93 & 0.058 & $-$19.9 ($gr$) & 3.2  & Y/-- & (7), (8)\\
\hline
SN~2012ht & Transitional & 1.39 & $-$18.40 & 0.003559 & $-$15.8 (Unf) & 4.3 & N/N
\ & (21)\\
iPTF13ebh & Transitional & 1.79 & $-$18.95 & 0.0133 & $-$14.5 ($R48$) & 5.9 & N/N & (22)\\
\hline
\textbf{PTF10ops} & Subluminous/02es-like & 1.12 & $-$17.77 & 0.062 & $-$15.6 ($R48$) & 3.2 & Y/-- & (23)\\
\textbf{iPTF14atg} & Subluminous/02es-like & 1.20 & $-$17.78 & 0.0213 & $-$18.5 ($R48$) & 3.8 & Y/Y & (24)\\
\hline
\textbf{MUSSES1604D} & Hybrid/He-det & 0.95 & $-$18.80 & 0.11737 & $-$19.6 ($g$) & 5.2 & Y/-- & (25)\\
\enddata
\tablecomments{\\
$^a$ Names of EExSNe Ia are denoted in boldface. \\
$^b$ The peak absolute magnitude in the $B$ band. Galactic extinction has been corrected for all golden early-phase SNe Ia. Reddening corrections of host galaxies are performed for SN~2010jn, SN~2012cg, SN~2013dy, SN~2015F, iPTF13ebh, iPTF16abc, and ASASSN-14lp by following related papers. For ASASSN-14lp, we tentatively adopt $R_V$ = 1.7 to give a reasonable peak magnitude for this spectroscopically normal SN Ia. \\
$^c$ The ``discovery phase" is the detection (the corresponding filter is shown in parentheses) time relative to the SN's $B$-band maximum in the rest frame. \\
$^d$ The ``photometric depth" is defined as a magnitude difference between the detection and the maximum in the same band. Due to the lack of unfiltered (Unf) peak magnitude of SN~2012fr and SN~2012ht, their photometric depths are calculated by using the $R$-band peak magnitude. \\
$^e$ (1) \citet{foley11}, (2) \citet{Hachinger13}, (3) \citet{nugent11}, (4) \citet{pereira13}, (5) \citet{zhang14}, (6) \citet{klotz12}, (7) \citet{firth14}, (8) \citet{walker15}, (9) \citet{ferretti16}, (10) \citet{zheng13}, (11) \citet{pan15}, (12) \citet{shappee16}, (13) \citet{cartier17}, (14) \citet{zheng17}, (15) \citet{richmond17}, (16) \citet{smitka15}, (17) \citet{marion16}, (18) \citet{miller18}, (19) \citet{hosseinzadeh17}, (20) \citet{cao16b}, (21) \citet{yamanaka12}, (22) \citet{hsiao15}, (23) \citet{maguire11}, (24) \citet{cao15}, (25) \citet{JJA2017}.\\
}

\end{deluxetable*}

\subsection{Early-phase normal and transitional SNe Ia} \label{subsec:Early-phase Normal and Transitional SNe Ia}

There are 12 normal SNe Ia in our golden sample. In general, their spectral features and luminosities well match those of normal SNe Ia. For three objects (SN~2010jn, SN~2012fr, and ASASSN-14lp) that show relatively high brightness but with normal-like spectral evolution, we classify them as normal SNe Ia at the bright end of the Phillips relation \citep{phillips93} by following previous studies \citep{Hachinger13,childress13,shappee16}. In addition to early optical observations, prompt $Swift$/UVOT follow-up observations were successfully triggered for six normal SNe Ia (\S3). However, none of the 12 normal SNe Ia show early light-curve excess in optical and UV wavelengths.

SN~2012ht and iPTF13ebh are categorized as ``transitional" type based on their fast-decline light curves and relatively high luminosities as opposed to subluminous/\ 91bg-like SNe Ia \citep{yamanaka12,hsiao15}. Although two transitional golden early-phase SNe Ia show quite different spectral features, smooth-rising light curves are found in both events at early phases.

\subsection{Early-phase luminous SNe Ia} \label{subsec:Early-phase Luminous SNe Ia}

Our golden early-phase SN Ia sample includes six SNe Ia with bright, slow-evolving light curves and shallow absorptions by intermediate-mass elements (IME), which are grouped into ``luminous/91T-like," ``luminous/99aa-like," and ``luminous/peculiar" subclasses in terms of their luminosities and specific spectral features. 

The relatively weak IME absorptions, prominent calcium and iron lines in the premaximum spectra of three luminous SNe Ia (SN~2012cg, iPTF16abc, and SN~2017cbv) are reminiscent of 99aa-like SNe Ia. Note that iPTF14bdn has been classified as a 99aa-like SN Ia by \citet{smitka15}. However, the early spectra dominated by iron-group elements (IGE), together with inconspicuous IME absorptions, indeed show a closer resemblance to those of 91T-like SNe Ia. Therefore, we classify iPTF14bdn as a 91T-like SN Ia in this paper. 

In recent years, several SNe Ia with excessively high luminosities and broad light curves have been discovered. Since the $^{56}$Ni mass required to produce such high luminosities is in conflict with that realized in the explosion of a 1.4 $M_{\odot}$ WD, they are frequently called ``super-Chandrasekhar-mass" SNe Ia (``super-$M_{\rm Ch}$" SNe Ia). Spectroscopically, super-$M_{\rm Ch}$ SNe Ia commonly show persistent C \textsc{ii} absorptions and moderately low photospheric velocities. Except for the unknown UV properties, optical photometry and spectroscopy of the golden early-phase SN Ia LSQ12gpw (\citealp{firth14,walker15}; spectra can be found in the Open Supernova Catalog\footnote{https://sne.space/}; \citealp{guillochon17}) indicate a high similarity to super-$M_{\rm Ch}$ SN Ia SN~2009dc \citep{silverman11}. We classify this 09dc-like super-$M_{\rm Ch}$ SN Ia into the luminous/peculiar subclass. Although the high peak luminosity, prominent IME absorptions, and especially the persistent carbon absorption of another golden luminous SN Ia, iPTF13asv---resemble super-$M_{\rm Ch}$ SNe Ia, its light curve evolves much faster than those of super-$M_{\rm Ch}$ and many 91T/99aa-like SNe Ia (Table 1). By following the main conclusion from \citet{cao16b}, we classify iPTF13asv as a luminous/peculiar SN Ia.

The UV/optical light-curve excess is found in the early phase of all 91T/99aa-like SNe Ia and LSQ12gpw (Figures 3--5), indicating an extremely high EEx fraction of luminous SNe Ia. Detailed information for all luminous EExSNe Ia is given in \S3.2.

\subsection{Early-phase subluminous SNe Ia} \label{subsec:Early-phase Subluminous SNe Ia}

The SN~1991bg-like, SN~2002cx-like, and SN~2002es-like SNe Ia are typical branches of subluminous SNe Ia. Mainly due to the fast-evolving light curves of 91bg-like SNe Ia, so far, none of them have been discovered at very early time. The 02cx-like SNe Ia occupy a large fraction of subluminous SNe Ia. Luminosities of 02cx-like SNe Ia show very large scatter, and their light curves evolve much slower than those of 91bg-like SNe Ia (see \citealt{foley13} for a review). Although several 02cx-like SNe Ia have been discovered over 10 days before reaching their $B$-band maxima, we are still missing the photometric information in the first few days after their explosions. The 02es-like SNe Ia are spectroscopically similar to 91bg-like SNe Ia, indicating a low temperature and ionization for both subclasses. The main characteristics of 02es-like SNe Ia are a moderately faint ($M_{B,max}$ $\sim$$-$17.7--$-$18.0) but slow-evolving (${\Delta}m_{15}(B)$ $\sim$1.1--1.3) light curve and lower ejecta velocity compared with those of 91bg-like SNe Ia \citep{ganeshalingam12,white15}. Currently, the number of reported 02es-like SNe Ia is small ($\lesssim$10), but two of them have been discovered at very early time. In total, only two 02es-like SNe Ia, PTF10ops and iPTF14atg, can be qualified as golden early-phase SNe Ia among published early-phase subluminous SNe Ia, and both of them show interesting early light-curve behavior (\S3.1). 

There is one early-phase supernova, iPTF14dpk, which was claimed as a 02es-like SN Ia by \citet{cao16a}. In \S3.4, we argue that the limited observational information cannot provide any crucial evidence for classifying iPTF14dpk as a 02es-like SN Ia; however, its peculiar early light-curve behavior indeed suggests that iPTF14dpk is more likely an early-phase core-collapse supernova (CCSN).

\subsection{Early-phase hybrid (He-det) SNe Ia} \label{subsec:Early-phase Hybrid (He-det) SNe Ia}

Hybrid SNe Ia are defined as SNe Ia that show normal-like light curves but with strong Ti \textsc{ii} absorptions commonly seen in subluminous SNe Ia at pre/around-maximum phase. \citet{JJA2017} recently discovered a prompt and red light-curve excess of a hybrid SN Ia, MUSSES1604D (SN~2016jhr), within $\sim$1 day of the explosion. The observed characteristics of MUSSES1604D, including the peculiar EEx and the Ti \textsc{ii} trough feature shown in this normal-brightness SN Ia, can be naturally explained by assuming that the SN is triggered by a precursory detonation of a thin He layer at the surface of a progenitor WD (the so-called ``He-shell-detonation" or ``He-det" scenario). Currently, MUSSES1604D is the only early-phase SN Ia belonging to this class.

\section{Early-excess SNe Ia} \label{sec:EEx SNe Ia}

The definition of EExSNe Ia is schematically illustrated in Figure 2. In general, early-excess features discovered so far can be visually classified into two types. The first type is a prominent and precursory EEx occurring a few days earlier than the $^{56}$Ni-powered main-body light curve (a ``spike-like" EEx), which can be easily distinguished (panels a \& b). The second type is a persistent (several days to more than 1 week) EEx that is coupled with the main-body light curve from the beginning (a ``bump-like" EEx), which can be distinguished when the early-excess feature is prominent (panel c) or the $^{56}$Ni-powered main-body light curve evolves quickly (panel d). In addition to two clear early-excess types, light curves influenced by an EEx scenario may not have clear spike- or bump-like features but show ``slow-rising" behavior in the early phase (the yellow region in panel e), which may originate from a weak bump that is covered by the main-body light curve, a weak excess observed in insensitive wavelengths or a relatively large survey cadence, and so on. Since such inconspicuous EEx will make the early light curve observationally broader, we hereafter classify these SNe as ``early-broad EExSNe Ia." Given that SN~2012cg shows the weakest early-excess feature in all previously reported EExSNe Ia\footnote{The UV light curves of SN~2012cg roughly meet the condition of panel d in Figure 2, and optical light curves are similar to the ``yellow+green" region in panel e; therefore, we use SN~2012cg as the reference for selecting early-broad EExSNe Ia.}, we compare our early SN sample with SN~2012cg to find out early-broad EExSNe Ia. Specifically, by shifting light curves of SNe Ia to match with that of SN~2012cg at peak, an SN Ia showing a broader rising-phase light curve and similar or narrow post-maximum light curve compared with that of SN~2012cg (the ``yellow+green" region in panel e) in any wavelength will be classified as an early-broad EExSN Ia. In other words, an SN Ia which shows a generally broader light curve than that of SN~2012cg but without a spike- or bump-like early light-curve feature (the ``yellow$+$red" region in panel e) or a fast-rising light curve (panel f) will be classified as a non-EExSN Ia. In our EExSN Ia sample (11 in total), nine of them show prominent early-excess features (Figures 3--5). The newly discovered SN Ia, SN~2017erp, and the previously reported SN~2012cg show relatively inconspicuous early-excess features. 

There are eight EExSNe Ia (iPTF14atg, iPTF14bdn, iPTF16abc, LSQ12gpw, MUSSES1604D, PTF10ops, SN~2012cg, and SN~2017cbv) in our golden sample. Since a more prominent early light-curve excess in UV/NUV wavelengths is predicted by various EEx scenarios, we further select three additional EExSNe Ia (SN~2011hr, SN~2015bq, and SN~2017erp) from 40 SNe Ia (14 of them are golden early-phase SNe Ia) that have UV/NUV light curves at relatively early phase. In addition, we introduce another two possible EExSN Ia candidates, ASASSN-15uh and SN~2015ak, in \S 3.2 but do not include them in our final sample due to incomplete information. Our final EExSN Ia sample includes seven luminous (91T/99aa-like \& LSQ12gpw), two subluminous (02es-like), one normal (SN~2017erp), and one He-det (MUSSES1604D) SNe Ia. Their basic information is shown in Table 2.

\begin{deluxetable*}{ccccccccc}
\tablenum{2}
\tablecaption{General Information on EExSNe Ia and Possible Candidates \label{tab:messier}}
\tablewidth{0pt}
\tablehead{
\colhead{Name} & \colhead{Subclass} & \colhead{${\Delta}m_{15}(B)$} & \colhead{Discovery Phase$^{a}$} & \colhead{Photometric Depth} & \colhead{EEx Shape$^{b}$} & \colhead{EEx Duration$^{c}$} & \colhead{Host Type}
}
\startdata
SN~2011hr & 91T-like & 0.92 & $-$13.6 ($B$) & 2.0 & c & Long & Sb (NGC2691) \\
SN~2015bq & 91T-like & $\sim$0.8 & $-$13.9 ($B$) & 2.1 & c & Long & -- \\
iPTF14bdn & 91T-like & 0.84 & $-$18.0 ($R48$) & 4.1 & c & Long & Irr (UGC8503) \\
SN~2012cg & 99aa-like & 0.86 & $-$17.1 ($B$) & 5.2 & d/e & Short & Sa (NGC4424) \\
iPTF16abc & 99aa-like & 0.95 & $-$17.8 ($g$) & 5.5 & c/d & Medium & Sb (NGC5221) \\
SN~2017cbv & 99aa-like & 1.06 & $-$18.9 ($B$) & 4.3 & c/d & Medium & SABc (NGC5643) \\
LSQ12gpw & 09dc-like & 0.76 & $-$19.9 ($gr$) & 3.2 & a/c & -- & -- \\
SN~2017erp & Normal & $\sim$0.9 & $-$17.3 (Unf) & ~3.6$^{d}$ & e & Medium & SABc (NGC5861) \\
PTF10ops & 02es-like & 1.12 & $-$15.6 ($R48$) & 3.2 & b/d/e & -- & -- \\
iPTF14atg & 02es-like & 1.20 & $-$18.5 ($R48$) & 3.8 & a & Medium & E (IC~831) \\
MUSSES1604D & He-det & 0.95 & $-$19.6 ($g$) & 5.2 & a & -- & S0-like \\
\hline
ASASSN-15uh & 99aa-like & $\sim$0.8 & $-$9.0 ($B$) & 1.1 & -- & -- & Sa-like \\
SN~2015ak & Normal & $\sim$1.1 & $-$15.1 (Unf) & ~2.2$^{d}$ & e & -- & Sb (ESO 108-21) \\
\enddata
\tablecomments{\\
$^{a}$ The ``discovery phase" is the detection (the corresponding filter is shown in parentheses) time relative to the SN's $B$-band maximum in the rest frame.\\
$^{b}$ The EEx shape is evaluated by using a single-band light curve, which includes the most complete early-phase information. If more than one photometric band meet the requirement, a light curve in the bluest wavelength is used. Letters correspond to the panels in Figure 2, which denote the possible EEx shape of each SN Ia.\\
$^{c}$ A rough estimation of the duration of early-excess features. Light curves used for the EEx shape evaluation and duration estimation are the same. ``Long," ``Medium," and ``Short" correspond to the EEx duration $T_{EEx}$ $\gtrsim$1 week, 0.5 week $\lesssim$ $T_{EEx}$ $\lesssim$ 1 week, and $T_{EEx}$ $\lesssim$ 0.5 week, respectively.\\
$^{d}$ Due to the lack of unfiltered peak magnitude of SN~2017erp and SN~2015ak, the photometric depths of two SNe are calculated by using the $Swift$/UVOT $U$-band light curves started 0.6 and 1.5 days after their discoveries, respectively.
}
\end{deluxetable*}

\subsection{02es-like and He-det EExSNe Ia} \label{subsec:02es-like and He-det EExSNe Ia}

The best-observed 02es-like SN Ia so far is iPTF14atg. The prominent early UV/optical excess of iPTF14atg can be basically explained by the companion-ejecta interaction scenario as proposed by \citet{cao15}. However, it is still unclear what kind of explosion mechanism under the SD progenitor scenario can explain the faint but slow-evolving light curve and spectral evolution of 02es-like SNe Ia. On the contrary, previous simulations by \citet{kromer16} suggest that the merger of two sub-Chandrasekhar-mass WDs could well reproduce the spectral evolution and light curves of iPTF14atg by assuming a low-metallicity progenitor system. They thus argued that the early light-curve excess of iPTF14atg cannot be attributed to the companion-interaction scenario.

The other 02es-like golden early-phase SN Ia, PTF10ops, also shows an interesting early-phase light curve. As shown in Figure 3, although only one data point of PTF10ops was taken at early time \citep{maguire11}, a very likely early light-curve excess can be distinguished in comparison to other golden early-phase SNe Ia and our companion-interaction model (\citealp{kutsuna15}; the cyan dashed curve in Figure 3), which implies that 02es-like SNe Ia may commonly show peculiar light-curve behavior in the early phase.

\citet{JJA2017} pointed out that the spike-like, red EEx of iPTF14atg is reminiscent of the He-det EExSN Ia MUSSES1604D, which originates from the supernova explosion triggered by the helium detonation at the surface of the progenitor. Given a significant amount of Ti production and the weak viewing angle effect predicted by the He-det scenario, the high EEx fraction (Table 3) and strong Ti \textsc{ii} absorptions shown in 02es-like SNe Ia can be explained with the He-det scenario. In addition, recent simulations also verify the possibility of reproducing the early light-curve excess of iPTF14atg with the He-det scenario (\citealp{maeda18}).

As the current sample size of early-phase 02es-like SNe Ia is still inadequate for statistical studies, it is uncertain that detections of EEx in both 02es-like SNe Ia is due to a viewing angle-independent EEx scenario for 02es-like SNe Ia or is just by chance.

\subsection{Luminous EExSNe Ia} \label{subsec:Luminous EExSNe Ia}

Surprisingly, all previously discovered early-phase 91T- and 99aa-like SNe Ia show early-excess features (Figures 3--5). Even though early-excess features in optical wavelengths are inconspicuous for some 91T/99aa-like SNe Ia, all of them show more prominent EEx in UV/NUV wavelengths. In particular, 91T-like SNe Ia seem to have stronger early-excess features than 99aa-like SNe Ia (Figures 4--6).

The early-phase $R$-band light curve of a luminous golden early-phase SN Ia, iPTF13asv, does not show the early-excess feature. Since both simulations from different EEx scenarios and previous observations suggest that observations in the $R$ band are not sensitive for detecting EEx, and the earliest photometry of iPTF13asv is about 1-2 days later than that of other luminous early-phase SNe Ia, we cannot confirm that the nondetection of EEx is due to the intrinsic distinction between iPTF13asv and other luminous EExSNe Ia or the imperfectness of early-phase observations for iPTF13asv. Here, we conservatively classify iPTF13asv as a non-EExSN Ia and do not make a comparison with other luminous EExSNe Ia in \S4.

Four luminous EExSNe Ia in our sample have been well investigated before. \citet{marion16} claimed that the EEx of SN~2012cg originates from the companion interaction in comparison to light curves predicted by \citet{kasen10}. Note that such a conclusion is controversial due to the lack of simulation-based comparisons with other EEx mechanisms. In the case of the early-excess feature of SN~2017cbv, the companion-interaction scenario is preferred based on the early-phase light-curve fitting result. However, other EEx scenarios indeed cannot be completely excluded, as discussed by \citet{hosseinzadeh17}. Other than SN~2012cg and SN~2017cbv, the EEx of iPTF16abc has been argued to be incompatible with the companion interaction based on an inconsistent light-curve fitting result and strong C \textsc{ii} absorptions at the early phase \citep{miller18}. \citet{smitka15} speculated that the blue EEx shown in a 91T-like SN Ia, iPTF14bdn, may originate from the radioactive decay in the outer-layer ejecta where the $^{56}$Ni abundance is greater than normal. However, there is no attempt to fit such a prominent early light-curve excess with any EEx scenario so far.

Two other luminous EExSNe Ia, SN~2011hr and SN~2015bq, and one possible luminous EExSN Ia candidate, ASASSN-15uh, are selected from published resources and the $Swift$ Optical/Ultraviolet Supernova Archive (SOUSA\footnote{http://swift.gsfc.nasa.gov/docs/swift/sne/swift\_sn.html. The reduction is based on that of \citet{brown09}, including subtraction of the host galaxy count rates and uses the revised UV zero points and time-dependent sensitivity from \citet{breeveld11}.}; \citealp{brown14}). SN~2011hr was classified as a luminous SN Ia which may bridge 91T-like SNe Ia and a super-$M_{\rm Ch}$ SN Ia, SN~2007if, in terms of high luminosity \citep{zhang16}. Although the expected $^{56}$Ni mass of SN~2011hr may be greater than the mean value of 91T-like SNe Ia, it can still be explained by the explosion of a Chandrasekhar-mass WD progenitor that experienced a more efficient detonation phase than normal SNe Ia. Additionally, the extremely high spectral similarity to SN~1991T from $-$13 days to 2 months after the $B$-band maximum indicates the physical resemblance between the two objects. In Figure 5, a prominent light-curve excess can be seen at $\sim$12 days before the $U$-band maximum. We therefore classify SN~2011hr as a luminous/91T-like EExSN Ia.

SN~2015bq was discovered by the intermediate Palomar Transient Factory (iPTF; \citealp{law09,rau09}) and well followed by $Swift$/UVOT from about 2 weeks before the $B$-band maximum. The high-luminosity, slow-evolving light curve, together with early-phase spectra composed by prominent IGE and inconspicuous IME absorptions, indicate that SN~2015bq is a typical 91T-like SN Ia. Interestingly, the prominent early light-curve excess of SN~2015bq shows very high similarity to two other 91T-like SNe Ia, iPTF14bdn and SN~2011hr, which indeed is much stronger than that of all 99aa-like EExSNe Ia. In \S4.2, we discuss the intrinsic connection between 91T-like and 99aa-like EExSNe Ia and the possible explosion mechanism of these luminous objects.

ASASSN-15uh was reported as a 98es-like SN Ia by the Asiago Transient Classification Program \citep{tomasella14,turatto15}. Due to the limited photometric and spectroscopic information of both the SN and its host galaxy, we took a low-resolution spectrum of the host galaxy with the Line Imager and Slit Spectrograph (LISS; \citealp{hashiba14}) mounted on the Nayuta 2-m telescope in December 2017. We derived a host redshift of about 0.0150 by using hydrogen emission lines, which is consistent with the reported redshift of the supernova ($\sim$0.0135; \citealp{turatto15}). High UV luminosity and slow-evolving light curve well match those of 91T/99aa-like SNe Ia. Spectroscopically, shallow IME absorptions together with prominent IGE absorptions around 10 days before the $B$-band maximum are also reminiscent of 99aa-like SNe Ia. For the early-phase light curve, a possible early-excess feature can be marginally distinguished from the earliest observation at $\sim$8 days before the $U$-band maximum (Figure 5). Since the early-phase information of ASASSN-15uh is insufficient to determine the rising behavior, we mark ASASSN-15uh as a possible EExSN Ia candidate and do not make a quantitative comparison with other luminous EExSNe Ia in the next section.

\subsection{SN~2017erp, the first normal EExSN Ia?} \label{subsec:SN2017erp, the first normal EExSN Ia?}

SN~2017erp was discovered at $\sim$17 days before its $B$-band maximum. Its early-phase spectral features, such as clear IME absorptions and high Si \textsc{ii} velocity, show a good similarity to those of a normal SN Ia, SN~2003W \citep{Jha17}. Particularly, early $Swift$/UVOT observations of SN~2017erp show interesting rising behavior in UV/optical wavelengths (Figures 3--5; see \citet{brown18} for more complete multiband light curves), which well meets our criteria of an early-broad EExSN Ia, especially for the $UVW1$-band light curve. Except for the peculiar early-phase light curve, a relatively slow-decline light curve toward the $B$-band peak magnitude of $\sim$$-$18.9 (Galactic extinction corrected) and the spectral evolution \citep{brown18} of SN~2017erp resemble those of normal SNe Ia, such as SN~2009ig and SN~2013dy. No matter what kind of mechanism accounts for the peculiar early-phase light curve of SN~2017erp, an early-broad EEx discovered in a normal SN Ia inspires us that the early light-curve excess can also be expected in a fraction of normal SNe Ia.

The normal-type EExSNe Ia may have been discovered even earlier. The prompt $Swift$/UVOT follow-up observation of SN~2015ak, a 99ee-like normal SN Ia identified by \citet{hosseinzadeh15}, shows a similar early-phase light-curve behavior to that of SN~2017erp, though the photometric uncertainty is much larger. The limited post-maximum information suggests that the light-curve evolution is normal-like and the $Swift$/UVOT $B$-band peak magnitude is $\sim$$-$18.5 mag by adopting the redshift of the reported host galaxy. Due to the uncertainties in both the SN classification and the EEx identification, we conservatively mark SN~2015ak as a likely normal EExSN Ia candidate.

\subsection{iPTF14dpk, a CCSN discovered at the shock-cooling emission phase?} \label{subsec:CCSN candidates at the shock-cooling emission phase}

\citet{cao16a} claimed that iPTF14dpk is a 02es-like SN Ia viewed from the direction that is far away from a nondegenerate companion star based on the nondetection of early light-curve excess and moderate similarities to iPTF14atg. We argue that the identification of this object is disputable because 1) a subluminous $R$-band light curve and limited spectral information\footnote{Spectra of iPTF14dpk were obtained at $-$10, $+$20, and $+$50 days after the $R$-band maximum, which are also similar to spectra of some SNe Ic at the same epochs, e.g. SN~2004aw \citep{taubenberger06}.} prevent us from identifying whether iPTF14dpk is a 02es-like, 02cx-like SN Ia or an SN Ic; 2) a starburst host galaxy is different from the hosts of all previously discovered 02es-like SNe Ia but commonly seen in 02cx-like SNe Ia, and also CCSNe where massive stars are actively formed; 3) according to the limiting magnitude of the last nondetection of iPTF14dpk, its $R$-band magnitude increased to $\sim$$-$17 with a rising speed over $-$1.8 mag night$^{-1}$ (Figure 3). Such a peculiar early light-curve behavior cannot be reproduced by any EEx mechanism of SNe Ia but resembles the cooling tail of shock breakout of CCSNe, the rising phase of which is too short to be caught by a day-cadence survey (e.g. SN~2006aj, \citealp{campana06}; SN~2011dh, \citealp{arcavi11}; iPTF15dtg, \citealp{taddia16}). 

Ambiguous classifications between subluminous SNe Ia and stripped-envelope CCSNe can happen sometimes. For example, although the spectral and light-curve resemblance between LSQ14efd and subluminous SNe Ia have been noticed by \citet{barbarino17}, they classified LSQ14efd as an SN Ic based on the early-phase light curve (Figure 3) and the incompatible Si \textsc{ii} velocity evolution to that of normal SNe Ia and a 02cx-like SN Ia, SN~2008A. We note that their argument about the velocity evolution can only exclude a part of SNe Ia but some subluminous SNe Ia (e.g. iPTF14atg) may have similar Si \textsc{ii} velocity evolution to that of LSQ14efd. However, given that both LSQ14efd and iPTF14dpk show unique early-excess features as opposed to other EExSNe Ia and the lack of crucial evidence for the SN classification, we argue that both objects are more likely CCSNe discovered at the shock-cooling phase.

\section{Discussions} \label{sec:Discussions}

\subsection{The fraction of EExSNe Ia and its implications} \label{subsec:The fraction of EExSNe Ia and its implications}

Current EEx scenarios can be classified into two different types in terms of the physical mechanism: interaction-induced (i.e. companion \& CSM interaction) and surface-radioactivity-induced (i.e. He-det \& ``surface-$^{56}$Ni-decay") scenarios. Recent simulations suggest that multiband photometric information during the EEx phase is crucial for differentiating EEx scenarios (e.g. the color evolution predicted by CSM interaction is generally faster than other scenarios, the companion-interaction may have relatively blue and slow color evolution at the EEx phase, etc.; see \citealp{maeda18} for details). However, the scatter of $^{56}$Ni-powered early-phase light curves \citep{firth14}, the viewing angle effect, and uncertainties of current simulations (especially in the UV wavelength) may prevent us from differentiating EEx scenarios in some cases. Therefore, further evidence from simulation-independent perspectives is particularly important. Since different EEx scenarios may not only give rise to different EEx fractions due to the viewing angle effect but also relate to specific SN Ia subclasses, statistically investigating the fraction of EExSNe Ia in each subclass can be a good starting point (Table 3).

\begin{deluxetable*}{ccccccc}
\tablenum{3}
\tablecaption{Supplementary Information on Early-phase SNe Ia in Different Subclasses and Their Early-excess Fractions\label{tab:messier}}
\tablewidth{0pt}
\tablehead{
\colhead{Subclass} & \colhead{Early-phase SNe Ia} & \colhead{EExSNe Ia} & \colhead{EEx Fraction} & \colhead{Observed Rising-} & \colhead{Photometric-depth} & \colhead{Host}
\\
\colhead{} & \colhead{Golden (Final$^{a}$)} & \colhead{Golden (Final)} & \colhead{Golden (Final)} & \colhead{time Limit$^{b}$} & \colhead{Limit$^{c}$} & \colhead{Type}
}
\startdata
Normal & 12 (17) & 0 (1) & 0 (5.9\%) & 18.2 (iPTF13dge) & 7.3 (SN~2011fe; $g$) & Various \\
Luminous (91T/99aa) & 4 (6) & 4 (6) & 100\% (100\%) & 18.9 (SN~2017cbv) & 5.5 (iPTF16abc; $g$) & Late-type\\
Luminous (peculiar) & 2 (1) & 1 (1) & 50\% (100\%) & 19.9 (LSQ12gpw) & 3.7 (iPTF13asv; $R48$) & Late-type\\
Transitional & 2 & 0 & 0 & 15.8 (SN~2012ht) & 5.9 (iPTF13ebh; $R48$) & Various\\
02es-like & 2 & 2 & 100\% & 18.5 (iPTF14atg) & 3.8 (iPTF14atg; $R48$) & Early-type\\
Hybrid & 1 & 1 & 100\% & 19.6 (MUSSES1604D) & 5.2 (MUSSES1604D; $g$) & Early-type\\
\enddata
\tablecomments{\\
$^{a}$ Since the earliest $R$-band photometry of iPTF13asv is about 1-2 days later than that of other luminous early-phase SNe Ia and there is no early UV/blue-optical photometry for this object, we exclude iPTF13asv from the final sample for the early-excess fraction calculation. Instead, we include five additional normal SNe Ia, SN~2007af, SN~2008hv, SN~2013gy, SN~2015ak, and SN~2017erp, and two 91T/99aa-like SNe Ia, SN~2011hr and SN~2015bq, which have the earliest UV/blue-optical information among other non-golden $Swift$/UVOT samples, in the final estimation.\\
$^b$ The ``observed rising-time limit" is the longest time interval (rest-frame) from the detection to the $B$-band maximum (the champion SN's name is shown in parentheses) in each SN Ia subclass.\\
$^c$ The ``photometric-depth limit" is the largest magnitude difference between the detection and the maximum in the same band (the champion SN's name and the corresponding filter are shown in parentheses) in each SN Ia subclass.\\
}
\end{deluxetable*}

Under the companion-interaction scenario, the early-excess feature can be observed only from specific viewing angles, which constrains the observable EExSNe Ia fraction to $\sim$10$\%$ \citep{kasen10}. Then, the probability of $\sim$10$^{-6}$ for a 100\% EEx detection of six 91T/99aa-like SNe Ia suggests that the early light-curve excess of 91T/99aa-like SNe Ia should be attributed to a different EEx scenario.

The CSM-ejecta interaction through the WD--WD merging channel is also difficult to explain EEx of 91T/99aa-like SNe Ia for the following reasons: i) short-lived carbon absorptions shown in 91T/99aa-like SNe Ia imply a low carbon-abundance environment; ii) such a high EEx fraction suggests a spherical distribution of CSM to eliminate the viewing angle effect, which may be difficult to realize because the CSM-interaction EEx requires a short lag time between the secondary WD disruption and the SN explosion (\citealp{raskin13,tanikawa15,levanon15}, \citeyear{levanon17}); iii) the long EEx durations of some luminous SNe Ia require a tremendous amount of CSM that is difficult to be produced by disrupting a companion WD \citep{maeda18}. Therefore, the detection of EEx in all 91T/99aa-like SNe Ia suggests the physical connection between EEx and the explosion mechanism (which may also be related to a specific progenitor system) of these luminous objects. 

Although EEx is always shown in 91T/99aa-like SNe Ia, a large scatter of the EEx strength can be found in Figures 4 and 5, which is barely related to the luminosity of main-body light curves. Instead, a likely correlation between the EEx strength and the equivalent width of the Si \textsc{ii} $\lambda$6355 line\footnote{In order to maximize the sample size of 91T/99aa-like EExSNe Ia for comparison, we use the magnitude variance from about $-$13 to $-$5 days after the peak epoch in the $UVW1$/$U$ band to describe the EEx strength. The smaller the magnitude variance, the stronger the early light-curve excess. Spectra used for the pseudo-equivalent width (pEW) calculation were taken around $-14$--$-13$ days after their $B$-band maxima.} is shown in Figure 6: a 91T/99aa-like SN Ia with a shallower Si \textsc{ii} $\lambda$6355 absorption has more prominent EEx. In other words, 91T-like SNe Ia generally have stronger early-excess features than 99aa-like SNe Ia. Such a correlation is independent of the peak luminosity (panels (c) and (d) in Figure 6).

Defining a ``non-EExSN Ia" is tricky, as it depends not only on the performed early-phase observations (cadence, imaging quality, wavelengths, etc.) but also on the strength of the EEx itself. By using the golden early-phase SNe Ia and other SNe Ia which have early-enough UV/NUV information, we try to present the currently ``best" non-EExSN Ia sample for the EEx fraction estimation. However, we note that even for some golden early-phase SNe Ia (e.g. SN~2010jn), it still remains a possibility that the non-EEx detection is due to the dimness of the EEx and/or unfavorable observing wavelengths (e.g. $V$ and $R$ band). In other words, the EEx fraction shown in this paper is a somewhat lower limit due to the probable incompleteness of EEx detections from previous observations. Therefore, an EEx fraction of 5.9\% (one of a total of 17 normal early-phase SNe Ia, which include 12 golden SNe Ia and five additional SNe Ia, SN~2007af, SN~2008hv, SN~2013gy, SN~2015ak, and SN~2017erp, that have the earliest UV/blue-optical photometric data among other non-golden normal SNe Ia\footnote{Given that some normal SNe Ia in our $Swift$/UVOT sample are not early enough for the non-EEx classification, we only include 12 golden normal SNe Ia and five additional normal SNe Ia, SN~2007af, SN~2008hv, SN~2013gy, SN~2015ak, and SN~2017erp, which have UV/blue-optical information at about 10 days before their maxima in the same band.}) or 11.8\% (by counting both SN~2017erp and SN~2015ak as normal EExSNe Ia) for normal SNe Ia cannot exclude any EEx scenario.

There were several attempts at searching for EExSNe Ia with datasets from large supernova survey projects \citep{hayden10,bianco11,ganeshalingam11}, but none of them reported a positive EEx detection. We argue that the previous studies are not in conflict with the high EEx fraction of 91T/99aa-like SNe Ia proposed in this paper for the following reasons. i) Since the purpose of those studies is to find significant early-excess features for testing the companion-interaction scenario (especially for the red-giant companion condition) by comparing their early-phase light curves with the model light curve predicted by \citet{kasen10}, they do not further investigate the ``small" early light-curve scatter in their plots. ii) Although only six 91T/99aa-like SNe Ia with very limited early-phase data points are included in \citeauthor{ganeshalingam11} (\citeyear{ganeshalingam11}; see Figure 3 therein), it is clearly shown that the large light-curve scatter at early phase is related to 91T/99aa-like SNe Ia. iii) The early light-curve scatter shown in Figure 1 of both \citet{hayden10} and \citet{bianco11} is similar to the scatter in Figure 3 of this paper, which may also relate to 91T/99aa-like SNe Ia in their samples. Since the sample composition in both papers is not clearly mentioned, further analysis of those samples may increase the total number of EExSNe Ia and improve the EEx-fraction estimation. Indeed, our finding about the correlation between early light-curve excess and 91T/99aa-like SNe Ia can be further supported by a very recent work by \citeauthor{stritzinger18} (\citeyear{stritzinger18}; accepted during the proof stage of this paper), which suggests that 91T/99aa-like SNe Ia tend to show very blue $B-V$ color evolution in the early phase. In addition to statistical studies, early-phase photometries of two $Kepler$-discovered SNe Ia, KSN~2012a and KSN~2012b, give good constraints on the non-EEx detection of two SNe \citep{olling15}, although we exclude them from the fraction calculation due to a lack of the spectroscopic identification. Given that both are more like normal SNe Ia in terms of the light-curve behavior, the non-EEx detections of two SNe should not influence the EEx fraction of 91T/99aa-like SNe Ia but may slightly decrease the EEx fraction of normal SNe Ia.

\subsection{From normal to super-$M_{\rm Ch}$ SNe Ia, the diversity and their intrinsic connections} \label{subsec: From normal to super-$M_{Ch}$ EExSNe Ia, the diversity and their intrinsic connections}

In order to explain the IGE-dominant absorptions in the early spectra of 91T/99aa-like SNe Ia, a more efficient detonation reaching the outermost region of the ejecta is frequently proposed (e.g. \citealp{zhang16}). This may account for the early light-curve enhancement due to high $^{56}$Ni production in the outer layer of the ejecta \citep{piro16}. However, as opposed to a spike-like EEx powered by a large amount of fast-decay radioactive elements from a precursory detonation of a thin He-shell predicted by the He-det scenario, a broadened, possibly bump-like early-phase light curve will be formed due to the radioactivity of substantial $^{56}$Ni at the outermost region of the ejecta. Such a prediction is indeed reminiscent of the early-rising behavior of luminous EExSNe Ia (Figures 4 \& 5).

Such an efficient detonation is expected in an extreme case of the delayed-detonation model, represented by the so-called gravitationally confined detonation (GCD; \citealp{plewa04,townsley07,jordan08,meakin09}). The buoyantly rising deflagration bubble ignited near the stellar core will finally trigger a detonation at the opposite pole from the breakout location of the deflagration bubble. Thanks to the mild expansion before the bubble reaches the surface, the detonation is more efficient than the traditional delayed-detonation scenario \citep{niemeyer97} and thus results in a larger amount of $^{56}$Ni extended to the ejecta's surface. According to \citet{jordan12}, the detonation happens at the antipodal point from where the star ejected the ash from its interior, giving rise to approximatly azimuthal symmetry upon the system. Therefore, the diversity of element distribution due to the viewing angle effect can be expected, e.g. more $^{56}$Ni will be generated at the surface on the opposite side of the off-center ignition point. Since the asymmetry of the detonation becomes smaller when getting to the inner region of the WD, the viewing angle effect in the earlier phase is more prominent. Such a prediction is basically in line with recent 3D simulations \citep{seitenzahl16}.

Although spectra produced by \citet{seitenzahl16} show clear distinctions from SN~1991T after taking into account the viewing angle effect, shallower-than-normal IME absorptions and luminous and slow-evolving light curves are qualitatively consistent with typical features of 99aa-like SNe Ia. Nevertheless, their simulation shows that only UV/NUV light curves and spectral features at wavelengths $\lesssim$ 5000 \AA~highly depend on the viewing angle, which suggests that a main spectral difference between 91T- and 99aa-like SNe Ia, the Si \textsc{ii} absorptions, cannot be completely attributed to the viewing angle effect through the GCD scenario.

Simulations of the GCD scenario also indicate that by switching the offset distance of the ignition point, the total $^{56}$Ni amount (and thus the EEx strength) and spectral features can be varied. In order to explain the extreme spectral features of 91T-like SNe Ia with the GCD scenario, an intuitive extrapolation is that 91T-like SNe Ia that have the strongest EEx and shallowest IME features correspond to the most extreme offset condition. By taking into account both the off-center level and the viewing angle effect, spectral differences between 91T- and 99aa-like SNe Ia, the correlation between the EEx strength and the equivalent width of the Si \textsc{ii} $\lambda$6355 line, and the luminosity scatter shown in 91T/99aa-like SNe Ia could be promisingly explained with the GCD scenario. In order to further testify to the GCD scenario and its relation to luminous SNe Ia, it is encouraged to investigate the upper limit of the initial offset and multiple-bubbles-triggered GCD models in future simulations. 

For the progenitor system of 91T/99aa-like SNe Ia, since the GCD-induced EEx can be realized through both single- and double-degenerate channels if the mass of merged WDs can reach the Chandrasekhar mass limit, early-excess features of 91T/99aa-like SNe Ia may not provide direct information on their progenitor systems.

A further question is whether both 91T/99aa-like and normal SNe Ia have the same origin. In principle, the luminosity and spectral features of normal SNe Ia can be reproduced by decreasing the offset distance of the ignition point so that the GCD scenario will finally get back to the traditional delayed-detonation scenario, which can also give rise to the smooth, small-scatter rising light curves that we see in normal SNe Ia. For the progenitor system of two SN Ia subclasses, nondetections of a surviving companion star and hydrogen emission lines in the nebular phase spectra of normal SNe Ia so far are in conflict with predictions by the SD progenitor scenario \citep{schaefer12,mattila05,leonard07,shappee13,lundqvist13,maguire16}, which is in opposition to the growing evidence of the possible connection between 91T/99aa-like SNe Ia and the SD progenitor scenario \citep{leloudas15}. Therefore, even with the same explosion physics (note that GCD in nature is a special case of the delayed-detonation scenario), normal and luminous SNe Ia may originate from different progenitor systems \citep{fisher15}.

If the amount of surface $^{56}$Ni is moderate compared with that of 91T/99aa-like SNe Ia, we may see an early-broad EExSN Ia instead of a bump-like one, which not only is in line with the early-phase light curve of SN~2017erp and previous simulations \citep{piro16} qualitatively but may also explain some ``slow rising," ``fast decline" light curves found in previous statistical studies \citep{hayden10,ganeshalingam11}. However, if the surface-$^{56}$Ni-decay scenario accounts for the EEx of the normal SN Ia SN~2017erp and the candidate SN~2015ak, a further question is why the $^{56}$Ni radiation through the traditional delayed-detonation scenario can realize such a large early light-curve scatter in normal SNe Ia. One possible explanation is that a small offset degree of the ignition point may also exist even under the traditional delayed-detonation scenario \citep{maeda10a}. In order to further understand the origin of such inconspicuous EEx shown in normal SNe Ia, a comprehensive investigation of SN~2017erp and SN~2015ak is highly recommended.

As we mentioned in \S2.2, early light-curve excess is also found in a 09dc-like SN Ia, LSQ12gpw, at $\sim$20 days before the $B$-band maximum. Although the explosion mechanism and progenitor system of super-$M_{\rm Ch}$ SNe Ia are still under debate, previous observations show evidence of dense CSM around super-$M_{\rm Ch}$ SNe Ia. \citet{yamanaka16} reported the near-infrared light echo from $\sim$40 to 110 days after the $B$-band maximum of a super-$M_{\rm Ch}$ SN Ia candidate, SN~2012dn, suggesting the presence of a wind from the pre-explosion system, possibly through the SD channel. If such a dense CSM can be formed nearer to the explosion site, not only the companion interaction but also the CSM interaction may explain the early light-curve excess of LSQ12gpw under the SD progenitor scenario.

On the other hand, \citet{taubenberger13} pointed out that the discrepancy between the extremely luminous peak and the faint light-curve tail of the super-$M_{\rm Ch}$ SN Ia SN~2009dc may require additional energy injection at early phase (see also \citealp{maeda09}), i.e. the interaction between SN ejecta and nearby CSM. Such confined CSM could be realized through the double-degenerate (DD) channel as the envelope predicted by the merger scenario would be ejected soon after the disruption of the secondary WD \citep{levanon15,tanikawa15}. Under the CSM-interaction scenario through the DD channel, not only the prominent early light-curve excess of LSQ12gpw but also the persistent C \textsc{ii} features that are commonly shown in super-$M_{\rm Ch}$ SNe Ia could be explained. Furthermore, although only one early-phase super-$M_{\rm Ch}$ SN Ia (i.e. LSQ12gpw) has been found, the discovery of EEx in this super-$M_{\rm Ch}$ SN Ia may imply an EEx scenario that is weakly affected by the viewing angle effect. Therefore, we speculate that the early light-curve excess in super-$M_{\rm Ch}$ SNe Ia is more likely powered by the CSM interaction through either the DD or the SD channel. Given that the CSM components are possibly dominated by different materials in two progenitor systems, spectroscopy at the EEx phase may provide new evidence of the progenitor system of super-$M_{\rm Ch}$ SNe Ia.

\subsection{The companion-interaction EExSNe Ia} \label{subsec: The companion-interaction EExSNe Ia}

The statistical study of EExSNe Ia not only further supports the multiple origins of previously discovered EEx but also reduces the possibility of the connection between luminous EExSNe Ia and the companion-interaction scenario to a large extent.\footnote{Note that if 91T/99aa-like SNe Ia truly originate from the SD progenitor system, it remains a possibility that the EEx of a small fraction of luminous SNe Ia may have multiple components.} However, as opposed to the bump-like EEx of 91T/99aa-like SNe Ia, a spike-like EEx of the subluminous EExSN Ia, iPTF14atg, cannot be interpreted by the $^{56}$Ni radioactivity from the outermost layer of the ejecta. Additionally, recent simulations by \citet{maeda18} also indicate a significant discrepancy between the early-phase light curve of iPTF14atg and their CSM-interaction models.

Although the EEx of iPTF14atg seems to be promisingly reproduced with the companion interaction \citep{cao15,maeda18}, whether its spectral evolution and general light-curve behavior can be reproduced through the SD channel is still under debate \citep{kromer16}. Alternatively, as we discussed in \S3.1 (also see \citealt{JJA2017,maeda18}), in terms of the spectral and EEx resemblance between MUSSES1604D and iPTF14atg, the He-det scenario is also promising for explaining 02es-like EExSNe Ia. As the current early-phase information of 02es-like SNe Ia is too limited to make a comparative analysis or give a stringent constraint from the statistical perspective, finding more early-phase 02es-like SNe Ia through ongoing transient surveys will shed light on the nature of this peculiar SN Ia subclass.

\section{Conclusions} \label{sec:Conclusions}

In this paper, we present general information on published, well-observed early-phase SNe Ia so far and summarize the characteristics of 11 (six reported and five unnoticed) EExSNe Ia in different subclasses. In particular, by investigating the connections and differences between these EExSNe Ia, new evidence of multiple origins of early light-curve excess and the implication for the explosion mechanism of 91T/99aa-like SNe Ia are presented.

We found that a 100\% EEx detection in six early-phase 91T/99aa-like SNe Ia is significantly in conflict with the prediction of the companion-ejecta interaction scenario but can be promisingly explained by the radioactive decay of a $^{56}$Ni-abundant outer layer or, alternatively (but less likely), interacting with spherically distributed CSM. In addition, the discovered correlation between the Si \textsc{ii} absorption feature and the strength of EEx also suggests an intrinsic connection between the explosion mechanism and the early light-curve excess of 91T/99aa-like SNe Ia. By investigating the early-excess behavior, post-EEx photometric/spectroscopic properties, and explosion models proposed for luminous SNe Ia, the surface-$^{56}$Ni-decay scenario is preferred for interpreting 91T/99aa-like EExSNe Ia. Specifically, we argue that the gravitationally confined detonation is a promising scenario for producing 91T/99aa-like SNe Ia with such a high EEx fraction. Also, spectral and early-excess differences between 91T- and 99aa-like SNe Ia and their intrinsic luminosity scatters could be qualitatively explained by taking into account the variation of the off-center distance of the initial ignition point and the viewing angle effect. In contrast to the high EEx fraction of luminous SNe Ia, early light-curve excess discovered in a normal SN Ia, SN~2017erp, and a possible candidate, SN~2015ak, so far suggests that EEx may accompany with a fraction of normal SNe Ia. Whether the EEx shown in normal SNe Ia is attributed to surface $^{56}$Ni decay through the traditional delayed-detonation mechanism requires further investigation.

Even though about a dozen of EExSNe Ia were successfully discovered in various SN Ia subclasses, we have not found any crucial evidence to support the companion-interaction scenario, even for the most promising candidate, iPTF14atg. The multiple origins of early light-curve excess suggest that EExSNe Ia may not be a superior indicator of the SD progenitor system as we originally expected, and we need to be more cautious when interpreting any newly discovered EExSNe Ia. Further understanding of the early-excess scenarios relies not only on individual studies of well-observed EExSNe Ia but also on systematical investigations of the early-phase light-curve behavior of each SN Ia subclass, which can be realized with ongoing survey projects such as the Zwicky Transient Facility (ZTF; \citealp{smith14}), the MUlti-band Subaru Survey for Early-phase SNe Ia (MUSSES; \citealt{JJA2017,miyazaki18}), and forthcoming transient surveys with the Tomo-e Gozen Camera mounted on the 1.05-m Kiso Schmidt telescope \citep{sako18} and the Large Synoptic Survey Telescope (LSST; \citealp{ivezic08}) in the near future.

\acknowledgments

We thank the anonymous referee for informative comments and suggestions. We also thank G. Hosseinzadeh, A. Miller, and M. Yamanaka for promptly providing access to published data of several SNe Ia in our sample. This work has been supported by Japan Society for the Promotion of Science (JSPS) KAKENHI Grants 18J12714 (J.J.), 18H04342 and 16H01087 (J.J. and M.D.), 18H05223 (J.J., M.D., K.M., and T.S.), 18H04585 and 17H02864 (K.M.), and 16H06341, 16K05287, and 15H02082 (T.S.).

\bibliographystyle{aasjournal}
\bibliography{Bibliography_EExSNeIa_Jiang}

\begin{figure}
\plotone{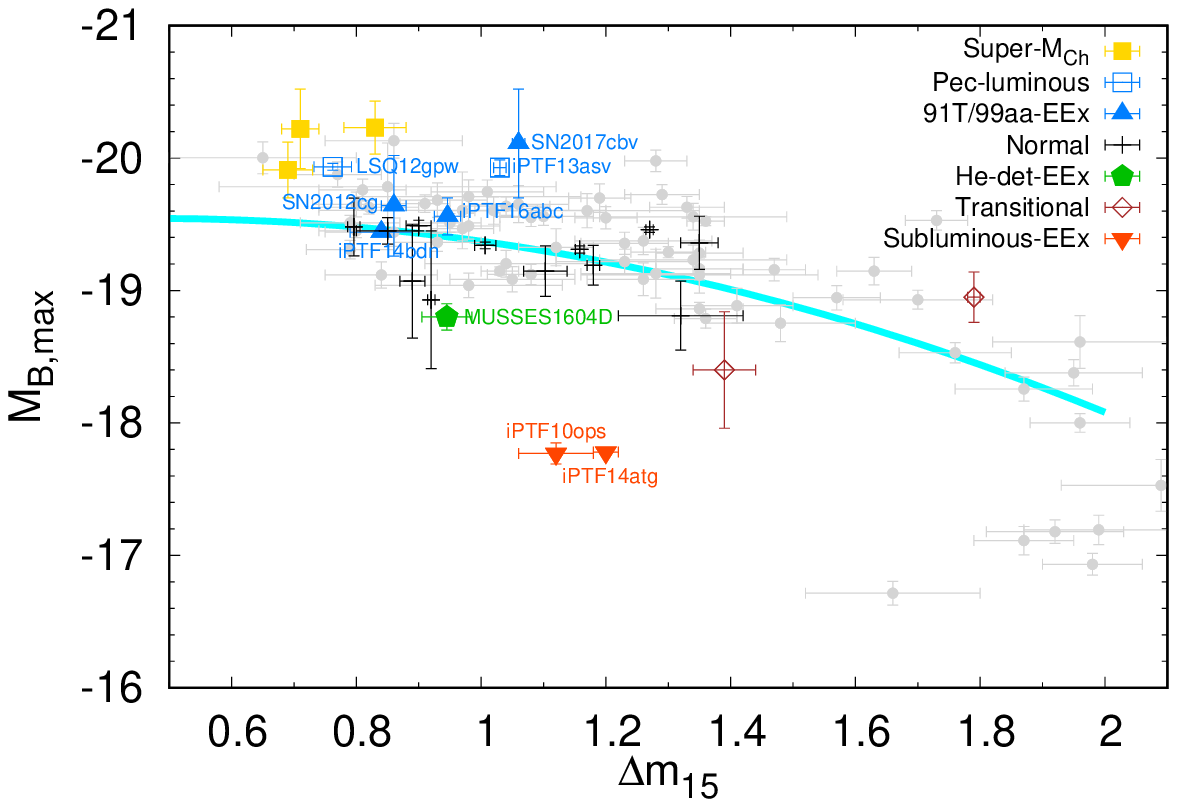}
\caption{Absolute $B$-band peak magnitude ($M_{B,max}$) vs. the light-curve decline rate, expressed by the magnitude decline in 15 days from the $B$-band peak ($\Delta$$m$$_{15}(B)$), for 23 golden early-phase SNe Ia. Four 91T/99aa-like (blue triangles), one luminous/peculiar (blue open square), two subluminous (red inverted triangles), and one He-det (green pentagon) SNe Ia show early-excess features. Gray points are from CfA3 samples in \citet{hicken09}. The Phillips relation is highlighted in cyan. \label{fig:fig1}}
\end{figure}

\begin{figure}
\plotone{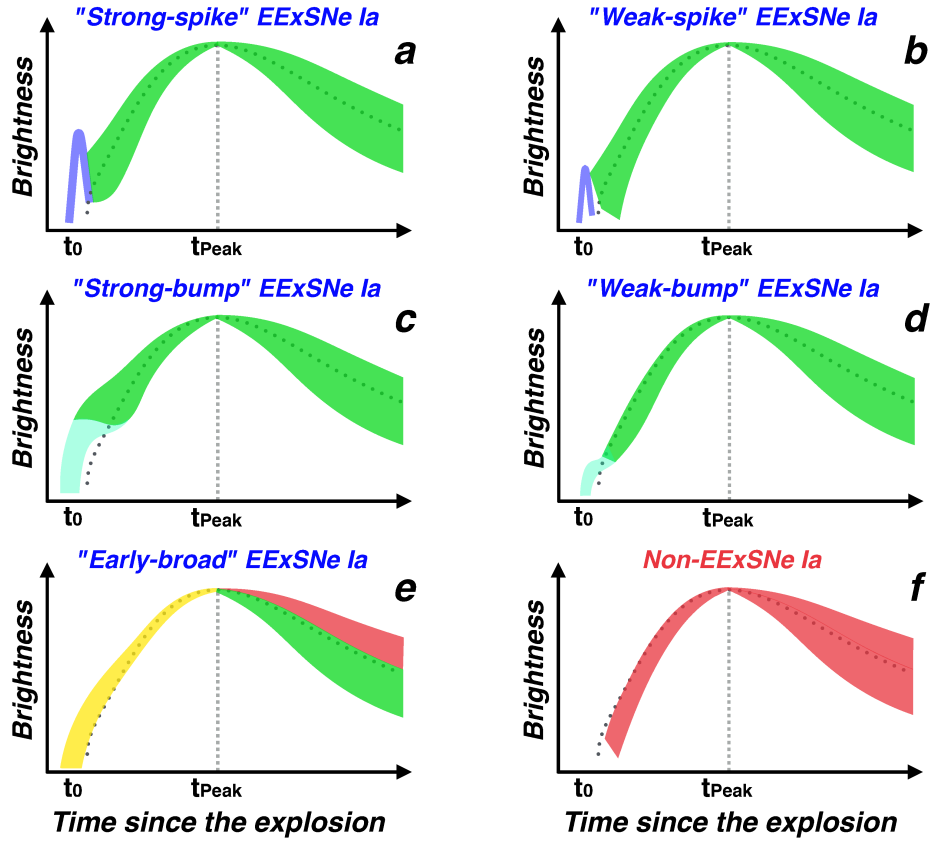}
\caption{Schematic diagram of different shapes of early-excess features. Spike-like (panels a \& b), prominent bump-like (panel c), and weak bump-like early-excess features but with a fast-evolving $^{56}$Ni-powered main-body light curve (panel d) can be easily distinguished. We further use SN~2012cg (a sketchy dotted light curve is shown in each panel, and all SN light curves are shifted to match with that of SN~2012cg at peak), which shows the weakest early-excess feature as the reference for selecting other early-broad EExSNe Ia. Specifically, an SN Ia showing a broader rising-phase light curve and similar or narrow post-maximum light curve compared with that of SN~2012cg (the ``yellow+green" region in panel e) in any wavelength will be classified as an early-broad EExSN Ia. SNe Ia with light curves in the ``yellow+red" region (panel e) and the red region (panel f) are classified as non-EExSNe Ia.\label{fig:fig2}}
\end{figure}

\begin{figure}
\plotone{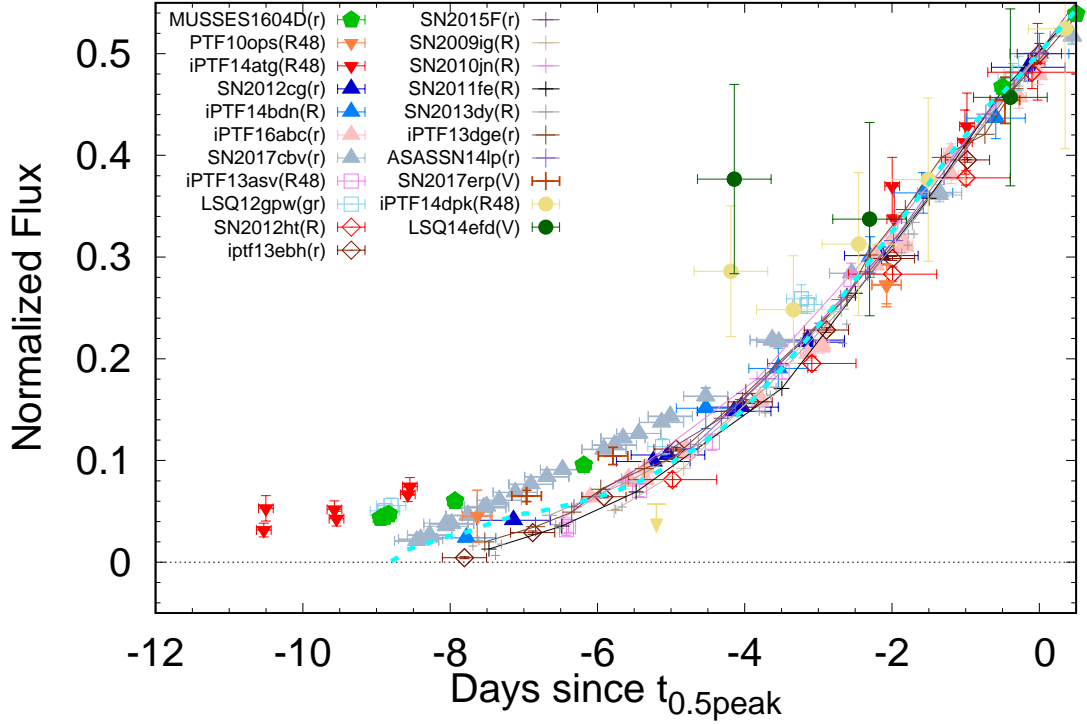}
\caption{Early-phase light curves of golden early-phase SNe Ia. Symbols are the same as in Figure 1. In order to compare early light-curve behaviors at a comparable wavelength coverage with the largest sample size, we select golden early-phase SNe Ia that have photometric information in either the PTF $R48$ band, Cousins $R$ band or SDSS $r$ band for comparison. Three normal SNe Ia with LSQ $gr$-band photometry are plotted with small gray crosses. Early light curves of SN~2012fr and SN~2016coj are not included due to the lack of early photometric information in a comparable wavelength coverage to other golden early-phase SNe Ia (note that neither of them shows early-excess feature even in blue wavelengths; \citealp{zhang14,zheng17}). As a reference of the early-excess feature of normal SNe Ia, we attached the early $Swift$/UVOT $V$-band light curve of the normal EExSN Ia, SN~2017erp. As can be seen, the early light-curve behavior of two cooling-emission CCSN candidates (filled circles), iPTF14dpk ($R48$) and LSQ14efd ($V$), are significantly different from those of golden early-phase SNe Ia. A cyan dashed curve is predicted by the interaction between SN ejecta and a 1.05 $M_{\odot}$ red-giant companion viewed from the companion side \protect\citep{kutsuna15}.\label{fig:fig3}}
\end{figure}

\begin{figure}
\plotone{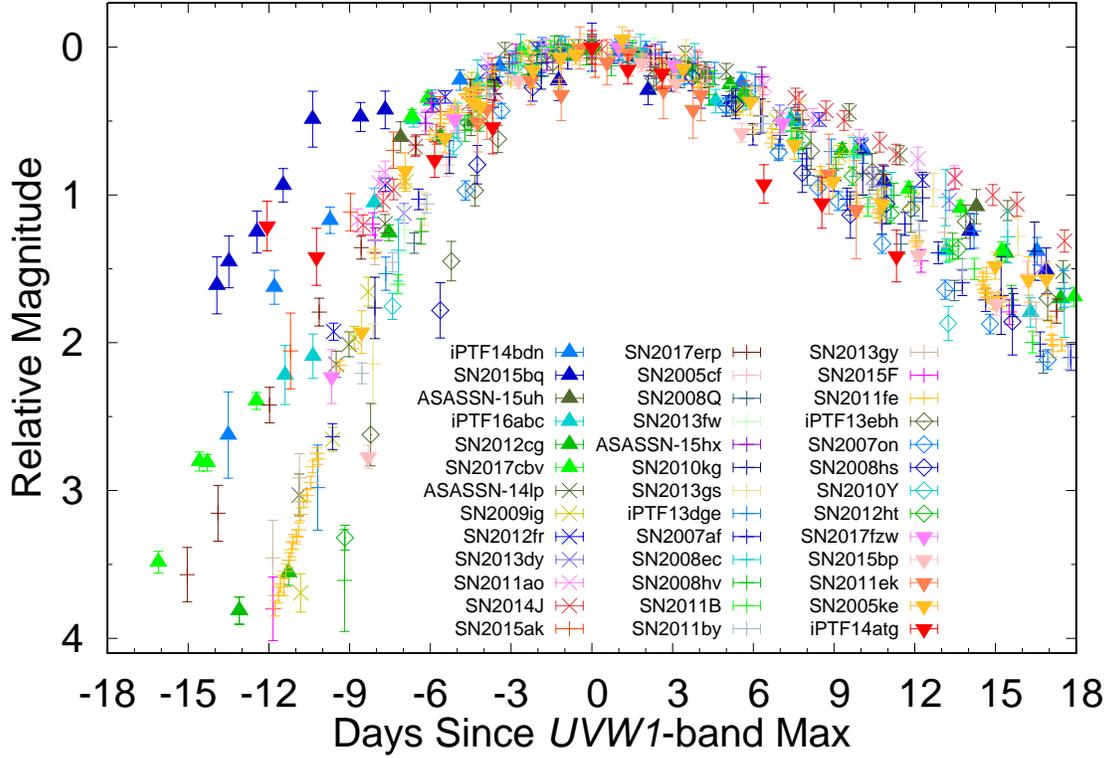}
\caption{Early-phase $UVW1$-band light curves of different SN Ia subclasses. Symbols are the same as in Figures 1 and 3 are used (diagonal crosses are specified for normal SNe Ia with slow-decline light curves). All light curves are shifted to the same peak magnitude for comparison. The 91T- and 99aa-like SNe Ia have clear excess compared to the rest, though with large scatter. In contrast, normal SNe Ia, except for SN~2017erp and SN~2015ak, show smooth, small-scatter, rising light curves though some of them (diagonal crosses) have similar post-maximum light curves to those of 91T/99aa-like SNe Ia. So far, we have not discovered any early-excess features from transitional, 91bg-like, and 02cx-like SNe Ia, which may be due to the extremely limited early-phase information of these SN Ia subclasses. \label{fig:fig4}}
\end{figure}

\begin{figure}
\plotone{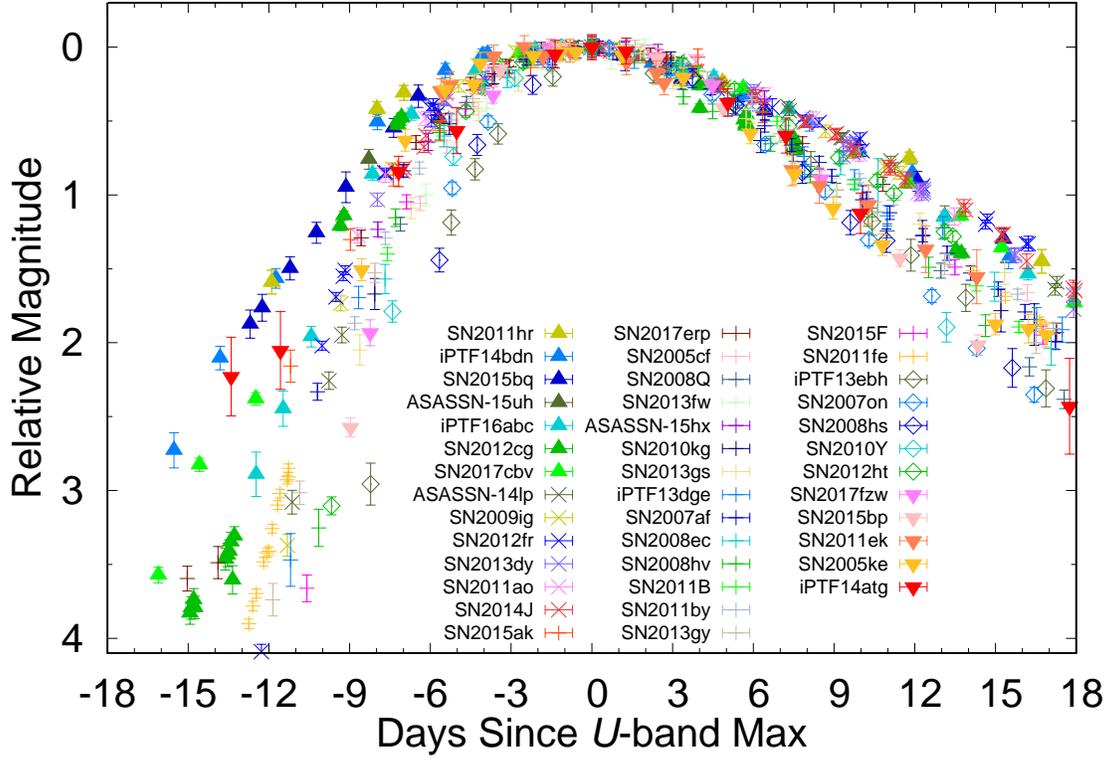}
\caption{Early-phase $U$-band light curves of different SN Ia subclasses. Symbols are the same as in Figure 4. All light curves are shifted to the same peak magnitude for comparison. Early light-curve excess can still be clearly discovered in 91T/99aa-like SNe Ia and a 02es-like SN Ia, iPTF14atg. Two normal SNe Ia, SN~2017erp and SN~2015ak, show marginal early-excess features in the $U$ band. The SNe Ia in other classes show smoothly rising light curves in the early phase. \label{fig:fig5}}
\end{figure}

\begin{figure}
\plotone{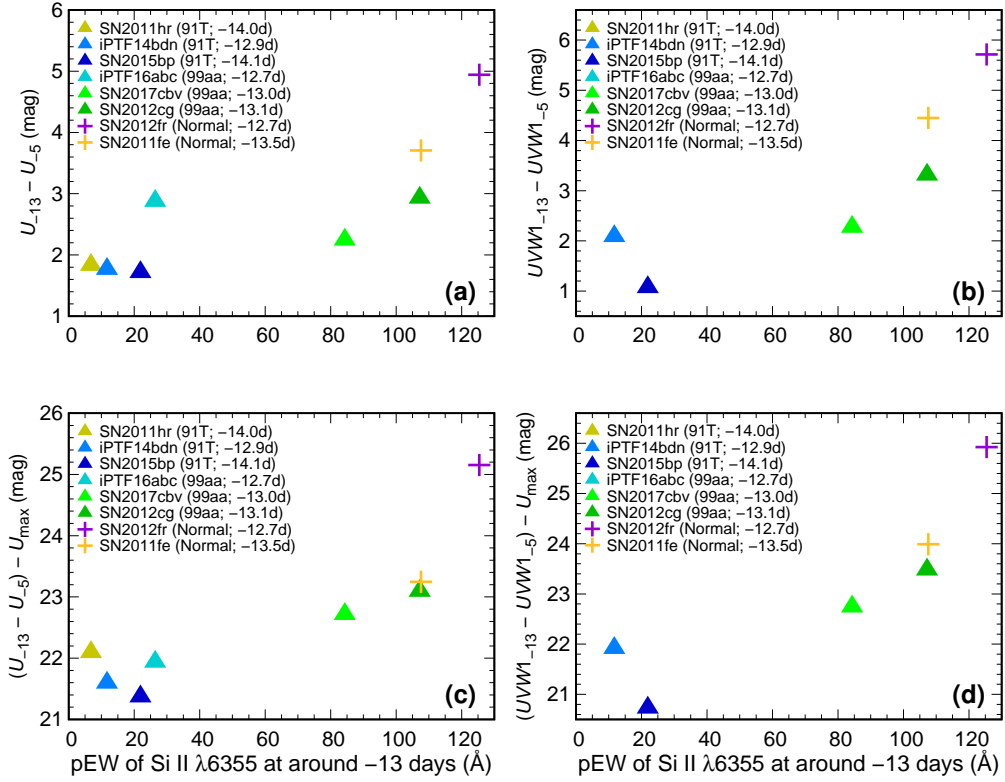}
\caption{Early $UVW1$/$U$-band excess vs. the pseudo-equivalent width (pEW) of the Si \textsc{ii} $\lambda$6355 line at about 13 days before the $B$-band peak of 91T/99aa-like EExSNe Ia. In order to maximize the sample size of 91T/99aa-like EExSNe Ia for the comparison, we use the magnitude variance from about $-$13 to $-$5 days after the peak epoch in the $UVW1$/$U$ band to describe the EEx strength (panels (a) and (b)). Two normal SNe Ia that have both photometric and spectroscopic information at a similar epoch are included as references. In panels (c) and (d), the correlation between the pEW of the early Si \textsc{ii} $\lambda$6355 line and the EEx strength is still clear after normalizing the magnitude variance to the $U$-band peak magnitude for each SN, suggesting that the proposed correlation is unlikely to be related to the SN's intrinsic brightness. The epoch (relative to the $B$-band maximum) of each spectrum used for the pEW calculation is shown in the legend. \label{fig:fig6}}
\end{figure}

\end{document}